\documentclass[letterpaper]{elsart}
\journal{Physica D}
\usepackage{natbib}\usepackage{graphicx}\usepackage{amsmath}\usepackage{amssymb}
\begin{document}
\begin{frontmatter}
\title{Quantum Chaos: Reduced Density Matrix Fluctuations in Coupled Systems}
\author[sn]{Sankhasubhra Nag} \ead{sankha@theory.saha.ernet.in}
\author[gg]{Gautam Ghosh} \ead{gghosh@theory.saha.ernet.in}
\author[al]{Avijit Lahiri\corauthref{cor}},\corauth[cor]{Corresponding author.}\ead{lahiri@cal2.vsnl.net.in}
\address[sn]{ Saha Institute of Nuclear Physics,1-AF,Bidhannagar,Kolkata 700 064,INDIA \thanksref{present}}
\thanks[present]{Present Address: Dept of Physics, Sarojini Naidu College for Women, Kolkata 700 028, INDIA}
\address[gg]{ Saha Institute of Nuclear Physics,1-AF,Bidhannagar,Kolkata 700 064,INDIA }
\address[al]{ Dept of Physics, Vidyasagar Evening College, Kolkata 700 006, INDIA}
\date{\today}
\begin{abstract}
Following a recent work (briefly reviewed below) we consider temporal fluctuations in the reduced density matrix elements for a coupled system involving a pair of kicked rotors as also one made up of a pair of Harper Hamiltonians. These dynamical fluctuations are found to constitute a reliable indicator of the degree of chaos in the quantum dynamics, and are related to stationary features like the eigenvalue and eigenvector distributions of the system under consideration. A brief comparison is made with the evolution of the reduced distribution function in the classical phase space.
\end{abstract}
\end{frontmatter}


\section{Introduction}
\label{sec:Introduction}
Quantum chaos deals, broadly speaking, with the study of quantum systems having chaotic classical counterparts. The problem of distinguishing  between the quantum behaviour of systems with chaotic and regular classical dynamics has been looked into for quite a long time and static features such as level repulsion and eigenvector distribution have been clearly identified as signatures of quantum chaos. However, while classical chaos essentially manifests itself in the dynamics of the system under consideration, analogous dynamical signatures of quantum chaos are not easily identified (see, e.g.,~\cite{Haake,Stockman,Casati-coll}). This is essentially because of the fact that  the quantum time evolution of a closed and bounded system must exhibit quasiperiodicity due to discreteness of the underlying  energy spectrum. The same quasiperiodicity can be seen in a periodically driven system due to the phenomenon of dynamical localisation~\cite{qloc,Fishman}. 

 Consequently, the search for dynamical features of quantum chaos has proceeded along different lines. In an early paper, Peres~\cite{peres1} pointed out that the quantum evolution of classically chaotic systems may show distinctive instabilities under external perturbations. A number of recent studies~\cite{fidelity1,fidelity2,fidelity3,ikeda}  suggest that such instabilities indeed constitute  a basic characteristic of such quantum  systems. Characteristic signatures in the time evolution of such systems are also to be found when these are coupled to other systems~\cite{sakagami,Sarben,nag}; the instabilities manifest themselves in the time series of the reduced density matrix (RDM) of the system under consideration. In all these studies, the distinctive quantum signatures are sought in the response of the system under consideration to external influence. 

 An alternative approach, to be pursued in the present paper, is to look within the system itself and to explore distinctive features of time evolution of its {\it subsystems} (an early work with a similar approach is to be found in
~\cite{graham}). Though an isolated closed system does not show any random features in time evolution, its subsystems do. Recent investigations reveal that the reduced entropy (or some entropy-like object measuring entanglement in subsystems) obtained from the RDM of a subsystem does show up distinctive features for systems possessing the stationary signatures of quantum chaos, as compared to the regular ones~\cite{AL,Arul1,Arul2,Arul3,ALN}. Thus, this entropy (and the analogous entanglement measure) may be looked upon as an indicator of quantum chaos. We shall present evidence in this paper that the decay of autocorrelations  in the temporal fluctuations of the RDM is strongly related with the lack of correlations among the Hamiltonian matrix elements of the system itself, and will relate this finding to the stationary features of quantum chaos. We will also present evidence indicating that the correlation among the elements of the Hamiltonian matrix  falls off as the corresponding classical motion becomes more and more chaotic(in this context, see~\cite{peres2}). It may be worthwhile to mention that such dynamical manifestations of quantum chaos, especially those  relating to the entanglement between subsystems, may be particularly relevant in the context of quantum information processing~\cite{qcomp,entang1,entang2,entang3,nielsen}.

 This paper is organised as follows. First, we outline (Sec.\ref{sec:BasicApproach}) our basic approach, following the papers \cite{AL,ALN}. In Sec.\ref{sec:TwoPrototypeSystems} we  follow \cite{ALN} in briefly presenting results for spin systems with two types of system Hamiltonians, one being described by a random matrix, and the other being a regular Harper's system (for a related model see~\cite{scott}). In the next section (Sec.\ref{sec:RealisticSubsystems}) we present results for subsystems of more realistic interacting systems. The following sections (Sec. \ref{sec:TheoreticalBackground} and  \ref{sec:HybridHamiltonians}) deal with a few other aspects of the problem, and concluding remarks (Sec. \ref{sec:ComparisonWithClassicalEntropy}).

\section{Basic Approach}
\label{sec:BasicApproach}

 The basic approach has already been discussed in an earlier paper \cite{ALN} (see also \cite{AL}). A pure state of a closed system continues to evolve unitarily as a pure state. Thus the von Neumann entropy defined in Eq.\eqref{eq:Svn} below (or the linear entropy defined in Eq.\eqref{eq:Sl}) remains constant irrespective of the nature of the Hamiltonian. However,  the reduced density matrix of a subsystem i.e. density matrix traced out over states of the complementary subsystem, does not evolve in a unitary or reversible way. This explains why the entropy of a system coupled to a heat bath increases with time and may show signatures of chaos \cite{nag}. As seen in \cite{nag}, the semi-classical entropy $S_{cl}=-\int\rho\ln\rho\;d p\;d q$  (where $\rho$ is the distribution function for an ensemble) calculated from the reduced Husimi distribution function for a subsystem also shows the rapid loss of correlation for a chaotic system (see below). 

 This idea serves as a clue for investigating the dynamical features
of quantum chaos for an isolated system. We proceed as follows.

If a system  $\bf{S}$ be a composition
of two subsystems ($\bf{A}$ and $\bf{B}$) , the density matrix of $\bf{S}$ may be traced out over the states of $\bf{B}$, giving the reduced density matrix (RDM) of $\bf{A}$ ( say $\rho_R$).The
system $\bf{S}$ evolves by the system Hamiltonian $H$
as \begin{equation}
\rho(t)=\exp(-iHt/\hbar)\rho(0)\exp(iHt/\hbar).\label{eq:evol}\end{equation}

\begin{figure*}
	\centering
		\includegraphics[height=6cm,width=9cm]{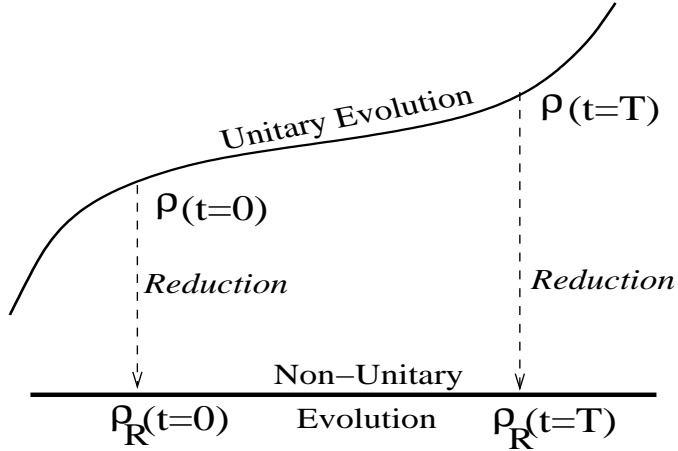}
	\caption{\label{cap:red2}Evolution of density matrix ($\rho$) and its reduced component ($\rho_R$).}
\end{figure*}

 Now, if $\rho(t)$ is traced out over $\bf{B}$ for each point of time to get the RDM
of $\bf{A}$ $\left(\rho_R(t)\right)$ at these successive time instants, then one can study the fluctuations in the matrix elements of $\rho_R(t)$ from the resulting time series. The schematic diagram for the time evolution of the RDM is given in Fig.\ref{cap:red2}.

One could study the fluctuations of the individual elements of the RDM by themselves. However, for the purpose of presentation it is best to confine ourselves to the following suitable variables viz. the von Neumann entropy($S_{VN}$) and linear entropy($S_{L}$) defined as
\begin{subequations}
\begin{eqnarray}
S_{VN} & = & -Tr(\rho_R\ln\rho_R)\label{eq:Svn}\\
S_{L} & = & 1-Tr(\rho_{R}^{2}).\label{eq:Sl}\end{eqnarray}
\end{subequations}

Both of them are indicative of the RDM elements fluctuations as a whole. Among the two, we choose the linear entropy for the presentation of most of our numerical results. Both these quantities are indicators of the degree of entanglement between the
states of $\bf{A}$ and $\bf{B}$.

\section{Two Prototype Systems}
\label{sec:TwoPrototypeSystems}
 We first consider two spin systems, once again by way of brief recapitulation of \cite{AL,ALN}. As a simple prototype
for quantum chaos we consider an $N\times N$ real symmetric Hamiltonian matrix $H_{c}$ with randomly distributed matrix elements for which the stationary features predicted by RMT are conformed to, and then we compare this with a second Hamiltonian matrix $H_{r}$(of the same dimension) that is equivalent to the Harper Hamiltonian on a torus (Eq.\ref{Harper}).
In terms of the co-ordinate ($q$) and momentum ($p$) on the torus,
the latter reads \begin{equation}
H_{r}={\gamma}_{1}\cos(2{\pi}p/P)+{\gamma}_{2}\cos(2{\pi}q/Q),\label{Harper}\end{equation}
 the torus being of area $PQ$ with $N$ basic states so that $\hbar=PQ/2{\pi}N$~\cite{leb, Dana}.

 Each of the systems may be considered as an assembly of $n$-interacting
spins such that $2^{n}=N$(see~\cite{scott}). With the Hamiltonian of the system $\bf{S}$
defined as above, we focus on a subsystem $\bf{A}$ made up of,
say, $p(<n)$ specified spins and look at the reduced density matrix
given by (using Eq.\ref{eq:evol}) \begin{equation}
{\rho_{R}}(t)={Tr}^{(n-p)}[\exp({\frac{-iHt}{\hbar}})\rho(0)\exp({\frac{iHt}{\hbar}})],\label{evolution}\end{equation}
 where $H=H_{c}$ or $H_{r}$ as the case may be, and where ${Tr}^{(n-p)}$
indicates partial trace with respect to states of the remaining $(n-p)$
number of spins. We then calculate $S_{L}^{(A)}(t)$ ($=1-Tr({{\rho}_{R}}^{2})$)
(or, alternatively, $S_{VN}^{(A)}(t)=-Tr(\rho_{R}\ln\rho_{R})$)
from Eq.(\ref{eq:Svn, eq:Sl}).

\begin{figure*}
\centering
\includegraphics[height=6cm,width=9cm]{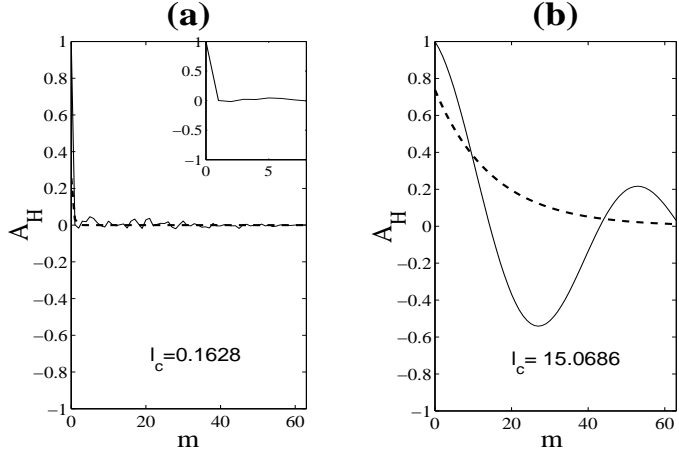}
\caption{\label{cap:AHspin}Autocorrelation of the Hamiltonian matrix as defined
in Eq.\ref{eq:autocor} for (a) $H_{c}$ (rapid initial fall is displayed in inset)  and (b) $H_{r}$ ($\gamma_1=\gamma_2=2.0$); (exponential fits are shown with dashed lines). }
\end{figure*}
 
\begin{figure}[ht]
 \centering
\includegraphics[height=6cm,width=9cm]{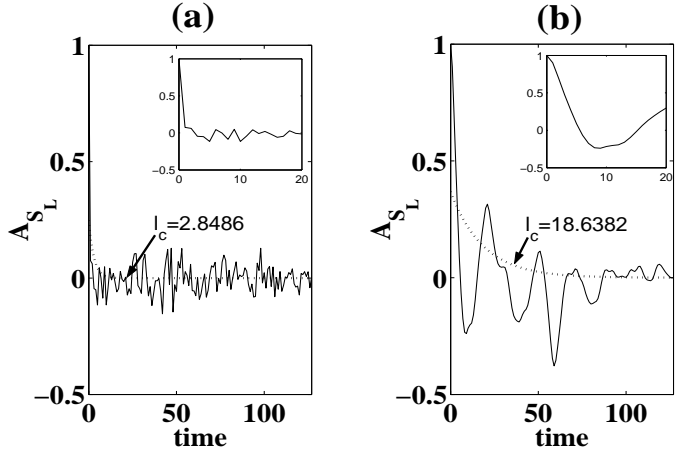}
\caption{\label{cap:AtrSpin}Autocorrelation of the time series of $S_L$ for spin systems when the system Hamiltonian is (a) $H_{c}$ and (b) $H_{r}$; $\hbar=0.592$ for both the cases (initial falls are shown on a shorter time scale in insets and exponential fits are given by dotted lines).}
\end{figure}

While ref. \cite{ALN} presents the numerical results and is to be referred to for details, figures \ref{cap:AHspin} and \ref{cap:AtrSpin} indicate graphically the distinctive characteristics of $H_c$ and $H_r$ for $n=7$, and $p=4$. Of these, Fig. \ref{cap:AHspin} (see caption) depicts the autocorrelation among matrix elements of $H_c$ and $H_r$. While the autocorrelation involves products like $H^{}_{k,l}H^{*}_{k+m,l+n}$, we look at the correlation in a direction `parallel' to the diagonal elements, namely
\begin{equation}
A_H(m)=\sum_{i,j}H_{i,j}H^{*}_{i+m,j+m},\label{eq:autocor}\end{equation}
 as being the most characteristic for the systems under consideration.

 In order to estimate the rate at which the autocorrelation falls off in a direction parallel to the principal diagonal, we fit the initial falling portion with an exponential fall and obtain the correlation length $l_c$. One finds from the figure that, as expected, the autocorrelation falls off much faster for $H_c$ as compared to $H_r$. Following ref~\cite{peres2} (see also \cite{Wien}), we suggested in ref.~\cite{ALN} that the degree of correlation among the Hamiltonian matrix elements can be looked upon as an indicator of the degree of regularity in the quantum dynamics. In the same work we also suggested that the dynamical manifestation of quantum chaos might be sought in the degree of randomness of the fluctuations of the RDM elements of subsystems of the system under consideration. The latter can be estimated from, say, the rate of fall of the autocorrelation of the linear entropy as a function of time, depicted in Fig.\ref{cap:AtrSpin} for $H_c$ and $H_r$. The typical time for fall of correlation is denoted with the same symbol, i.e., $l_c$, and one finds that there is indeed a good correspondence between the divergences in the values of $l_c$ for the Hamiltonian matrix and for the temporal fall of correlation of the RDM elements.

To summarise, it appears worthwhile to look for a reliable indicator of quantum chaos in the temporal fluctuations of RDM elements of subsystems. These fluctuations are of a random nature for chaotic systems as compared to regular ones, as borne out by the `toy Hamiltonians' $H_c$ and $H_r$. The degree of this randomness reflects, in a way, the lack of correlation among the elements of the Hamiltonian matrix of the system under consideration. Other measures for estimating the degree of randomness in the temporal evolution of the RDM are to be found in \cite{ALN}.

\section{Coupled Kicked Rotors and Coupled Harper Systems}
\label{sec:RealisticSubsystems}

We now turn our attention to systems made up of more realistic subsystems in order to see if the observations made in ref. \cite{ALN} and briefly outlined above can be claimed to have general validity. 

 First, we consider below a system made up of two kicked rotors coupled together. The next subsection will deal with a system composed of two coupled Harper Hamiltonians.

\subsection{Coupled kicked rotors}
\label{sec:CoupledKickedRotors}

 Kicked rotors are widely referred to in the context of quantum chaos (see e.g. \cite{qloc,Izr}). A single kicked rotor is described by the Hamiltonian
\begin{equation}
H=\frac{p^{2}}{2}+K\cos q\sum_{n}\delta(t-n\tau),\label{eq:krotor}\end{equation}
 where $K$ is the kick strength and $\tau$ is the interval between two consecutive
kicks. The classical dynamics is described conveniently in terms of the standard map, for which chaotic features are known to dominate for $K>K_{c}\sim.97$ (for $\tau=1$).
\begin{figure}[ht]
\centering
\includegraphics[height=6cm,width=9cm]{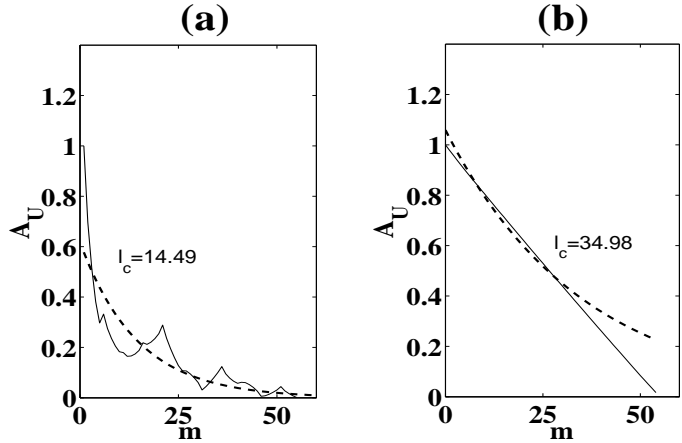}
\caption{\label{cap:AUkRot}Autocorrelation $A_U$ among evolution matrix elements defined in Eq.\ref{eq:Uautocor} for the coupled kicked rotor system; kick strengths
are  (a) $K=10$ and (b) $K=0.1$ (see text); $\tau=1.0, c_r=2.0$ for both cases; corresponding correlation lengths $l_c$ are shown in the figure. }
\end{figure}

 We consider the composite system made up of two kicked rotors and described by the Hamiltonian
\begin{eqnarray}
H \;& \!\!= &\; \!\!\frac{p_{1}^{2}}{2}+K_{1}\cos q_{1}\sum_{n}\delta(t-n\tau)+\frac{p_{2}^{2}}{2}\nonumber \\
 &  & \!\!\!\!\!\!\!\!+K_{2}\cos q_{2}\sum_{n}\delta(t-n\tau)+c_r\sin q_{1}\sin q_{2},\label{eq:coupkrot}\end{eqnarray}

 where $c_r$ is the coupling strength. Denoting the RDM for one of the systems
by $\rho_{R}$, we look at the autocorrelation of the time series for
$S_{VN}=-Tr\left(\rho_{R}\ln\left(\rho_{R}\right)\right)$ for various values
of $K_{1}=K_{2}=K$(say). 
\vskip .5 cm
 The single step evolution operator $U$ is a complex two-dimensional unitary array, and the autocorrelation of the elements of this array (we consider a finite dimensional truncation) can be expressed in terms of 
\begin{equation}
A_U(m)=\sum_{k,l}\left|U_{k,l}\right|\left|U_{k+m,l+m}\right|.\label{eq:Uautocor}\end{equation}
 where we have once again limited ourselves to correlation in a direction parallel to the principal diagonal (as in Eq.~\eqref{eq:autocor}) and have considered only the modulii of the matrix elements, ignoring the phases (the latter lead to similar results).

  Fig.~\ref{cap:AUkRot}(a,b) depict the variation of $A_U(m)$ with $m$ for two values of the non-linearity parameter $K$ (see caption), and for the coupling strength $c_r=2.0$ . The choice of parameters  is such that for Fig.\ref{cap:AUkRot}(a) the classical phase space of each rotor as also the phase space of the composite system is dominated by chaotic orbits, while for Fig.\ref{cap:AUkRot}(b) the corresponding phase spaces are predominantly regular (note that for small values of both $K_1$ and $K_2$, the phase spaces are regular regardless of $c_r$, since the Hamiltonian can then be effectively written as the sum of two uncoupled free rotors). One finds a close correspondence between the classical phase space and the correlation among the matrix elements of the single step evolution operator (the basis chosen is the direct product of the momentum states of the individual rotors).

\begin{figure}[ht]
\centering
\includegraphics[height=6cm,width=9cm]{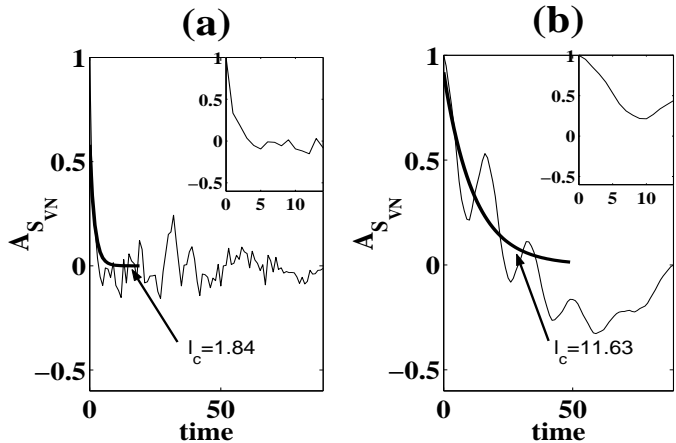}
\caption{\label{cap:AtrkRot}Autocorrelation of the time series for $S_{VN}$ (see text)
taking one of the coupled kicked rotors as subsystem where parameters as in Fig.\ref{cap:AUkRot}  (exponential fits in thick lines).}
\end{figure}

\begin{figure}[ht]
	\centering
		\includegraphics[height=6cm,width=9cm]{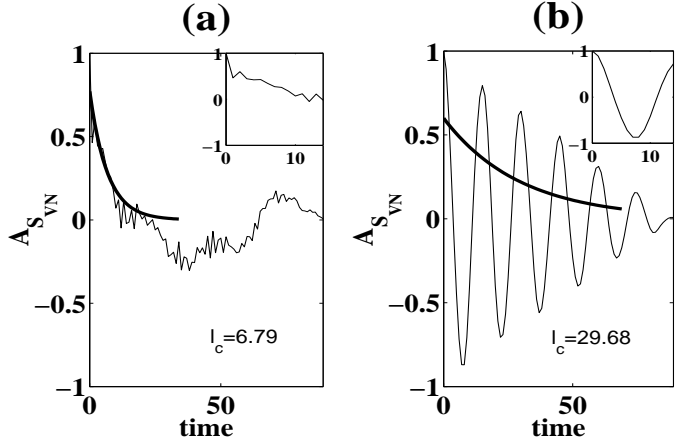}
	\caption{\label{cap:AtrkRotSp} Temporal variation of autocorrelation of \(S_{VN}\) when (a) \(K_1=0\) (\(K_2=10\) and (b) \(K_1=K_2=0\) ($c=2.0$ for both the cases; initial fluctuations in shorter time scale are in the insets, the fits are thick lines in the main windows ).}	
\end{figure}
 Fig.\ref{cap:AtrkRot}(a),(b)  depicts the temporal fluctuation of the RDM (obtained by taking partial trace over the rotor with variables ($q_2,~p_2$)), expressed through the variation of autocorrelation of $S_{VN}$ as indicated above (parameter values are the same as in Fig.\ref{cap:AUkRot}(a), (b) respectively). One finds a close correspondence between the fall of temporal autocorrelation in $S_{VN}$ (similar results are obtained for $S_L$ [not shown here]) obtained for a subsystem and the lack of correlation among matrix elements of the single step evolution operator $U$ of the composite system. This once again goes to show that the temporal fluctuations in the RDM elements can indeed be looked upon as an indicator of quantum chaos.

 It is important to note that the randomness of the reduced density matrix fluctuations is a property of the (composite) system under consideration, and {\it not} of the subsystem whose RDM we are looking at. In other words, the nature of RDM fluctuations depends on whether or not the system as a whole is chaotic, and is random for any {\it arbitrary } choice of the subsystem. This is seen from Fig.\ref{cap:AtrkRotSp} (a), (b), where, in (a) we take $K_1=0$, i.e., the first rotor, for which the RDM is constructed, is regular, while $K_2$ and $c_r$ are given values such that the entire system is predominantly chaotic. We compare this with (b) where we take $K_1=K_2=0$, so that the entire system, as well as both subsystems, is regular. The rapid loss of autocorrelation in (a) compared to the quasiperiodic fluctuations in (b) tells us that the RDM fluctuations are indeed characteristic of the system under consideration, and not of the subsystem one is looking at.

\subsection{Coupled Harper systems}
\label{sec:CoupledHarperSystems}

 In continuation of the observation made in the last paragraph, we now consider a system consisting of a pair of Harper Hamiltonians coupled together. Here each individual subsystem is regular, while the nature of the classical phase space of the coupled system is determined by the strength of coupling. 

 The phase space for each subsystem is periodic in both $q$ and $p$ with periods, say, $Q$ and $P$ respectively. In other words, the phase space can be taken as a toroid with dimensions $Q$, $P$. The quantum mechanics on a toroidal phase space imposes an additional condition \(PQ/2\pi\hbar=N\), where $N$ must be an integer (see \cite{leb, Dana}).  

 We take the coupled system as 
\begin{equation}
H=H_{1}+H_{2}+c_{h}\sin q_{1}\sin q_{2}\label{eq:coupHar}\end{equation}
where $H_{1}$and $H_{2}$ are two Harper Hamiltonians (Eq. \ref{Harper})
with identical parameter values and $c_{h}$ is the coupling constant. The form of the interaction has been chosen such that the phase space of the composite system is a $4D$ torus while the quantisation condition is to be separately imposed for each set of $q$ and $p$.

\begin{figure}[ht]
\centering
\includegraphics[height=6cm,width=9cm]{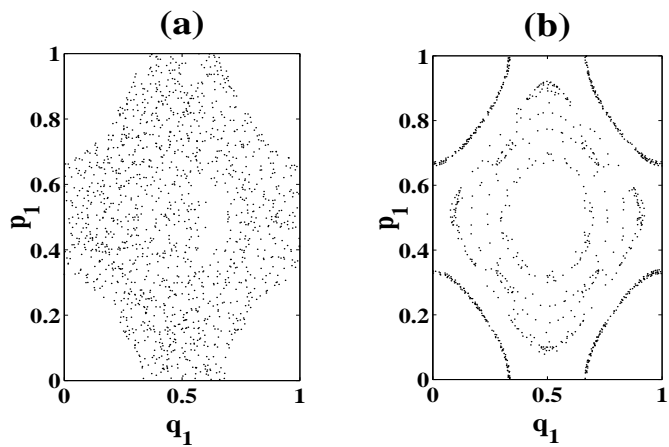}
\caption{\label{cap:Poincare}Poincare section for the classical dynamics
of the coupled Harper system when the coupling strength is (a)$10$
and (b)$0.1$ ($L_{1}=L_{2}=K_{1}=K_{2}=2$ for both the cases).}
\end{figure}

\begin{figure*}
\centering
\includegraphics[height=6cm,width=9cm]{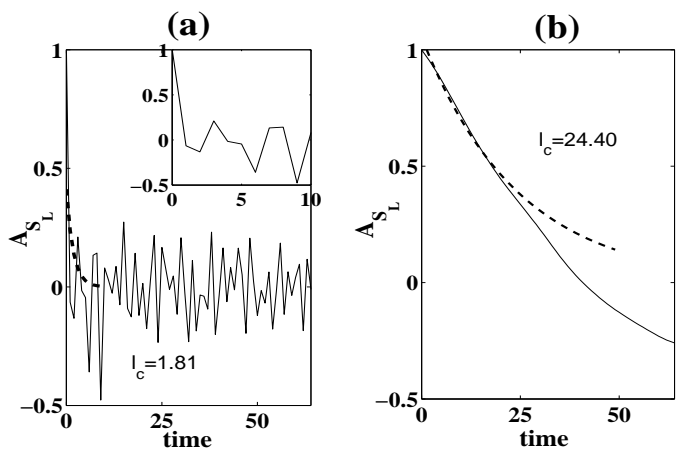}
\caption{\label{cap:AtrHar}Autocorrelation of the time series of $S_L(=1-Tr(\rho_{R}^{2}))$
for one of the coupled Harper systems when the coupling
strengths are as mentioned in Fig.\ref{cap:Poincare}; $\hbar=0.628$.}
\end{figure*}
\begin{figure*}
\centering
\includegraphics[height=6cm,width=9cm]{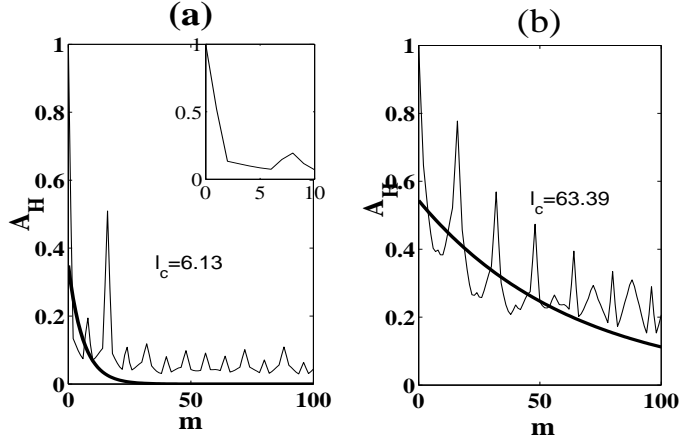}
\caption{\label{cap:AHhar}Autocorrelation of the Hamiltonian matrix as defined
in Eq.\ref{eq:autocor} for the coupled Harper systems for
coupling strengths as mentioned in caption of Fig.\ref{cap:Poincare}.}
\end{figure*}

 Fig.\ref{cap:Poincare} shows the classical phase space in a Poincare section along $q_1$-$p_1$ where one finds that the degree of chaos in the composite system can be tuned through the coupling strength $c_h$. Fig.\ref{cap:AtrHar}, on the other hand, gives the time series for autocorrelation of the RDM elements, expressed through the linear entropy, for the same two values of the coupling strength. One finds a close correspondence between the two sets of data which show that a classically regular (resp. chaotic) phase space indeed results, in the quantum context,  in a quasiperiodic (resp. random) variation in the RDM elements. 
 Additionally, Fig.\ref{cap:AHhar} depicts the correlation among the Hamiltonian matrix elements, once again for the same two values of the coupling strength, and indicates, as already observed, a close correspondence between the degree of correlation among these matrix elements and the nature of quantum dynamics as revealed through temporal fluctuations in RDM elements.

\begin{figure}[h]
\centering
\includegraphics[height=6cm,width=9cm]{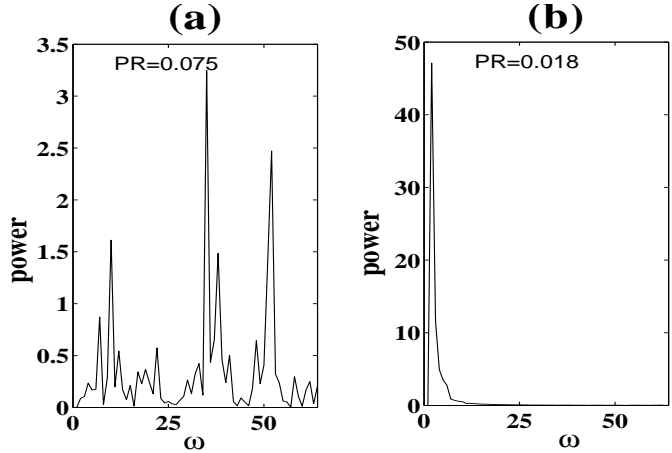}
\caption{\label{cap:PwHar}Power spectra of fluctuations in the time series
of $Tr(\rho_{R}^{2})$ for the coupled Harper systems with
coupling strengths as mentioned in caption of Fig.\ref{cap:Poincare}.}
\end{figure}

 The degree of randomness in the RDM fluctuations is also revealed in the power spectrum of the time series for, say, $Tr(\rho_{R}^{2})$, which is simply the Fourier transform of the corresponding temporal autocorrelation. A random fluctuation is indicated by a broad-band power spectrum while a more regular temporal variation corresponds to a power spectrum with a few localised peaks. A convenient way to quantify this \cite{AL,ALN} is to compute the participation ratio (PR) for the power spectrum. Fig.\ref{cap:PwHar} depicts the power spectrum for the coupled Harper model with the same parameter values as in Fig.\ref{cap:AtrHar} and \ref{cap:AHhar}, and once again confirms our basic observation that the RDM fluctuations are a reliable indicator of the degree of quantum chaos charactering the system under consideration.

\section{RDM Fluctuations related to Eigenvalue and Eigenvector Statistics}
\label{sec:TheoreticalBackground}
 The evidence presented above seems to indicate that the degree of correlation among the Hamiltonian matrix elements of a classically chaotic system is characteristically small compared to that for a regular one (which is already implied in the random matrix theory (see \cite{peres2, Wien}) and corresponds to the stationary features related to eigenvalue and eigenvector distributions for chaotic systems), and this, in turn is correlated with the degree of randomness in the temporal fluctuations of RDM elements of subsystems. We indicate below how this correlation comes about.

 Meanwhile we take note of the fact that, in the classical situation, there exists a whole spectrum of the degree of randomness in the underlying dynamics, ranging from fully regular motion, through `soft chaos', to motion characterised by `hard chaos'. One can be more specific and quantify the degree of chaos through such indicators  as the measure in phase space of the region occupied by chaotic trajectories, and the average decay time of autocorrelations. Our numerical results presented above indicate that a similar spectrum of the degree of chaos can also be identified in the quantum regime through measures related to the RDM fluctuations (see \cite{ALN} for evidence based on a model Hamiltonian with a parameter effecting a convenient tuning for chaos), as also through ones related to stationary features such as the nearest neighbour level statistics (e.g., the Brody distribution that interpolates between the Wigner and Poisson distributions \cite{Haake,Stockman,brody}). Our analysis below tends to confirm these observations.

 The reduced density matrix elements are given by

\[
\left(\rho_{R}\right)_{mn}=\sum_{l}\rho_{ml,nl}\]
where $\rho$ is written in the usual block matrix form, and $\rho_{mk,nl}$ is the matrix element in the $k$'th row and
$l$'th column of the $m$-$n$'th block of the matrix. In all our computations we have started with an initial composite density matrix which is a direct product of density matrices for the subsystems chosen :

\[
\rho=\rho_{A}\otimes\rho_{B}.\]

 Now, according to Eq.\ref{evolution},

\begin{eqnarray*}
\left(\rho_{R}\right)_{m,n} & = & \sum_{l}\rho_{ml,nl}\\
 & = & \sum_{a,b}\exp\left(-\frac{i(E_{a}-E_{b})t}{\hbar}\right)\times \\
 & & \sum_{l}\left\langle m,l\right|\left.E_{a}\right\rangle \left\langle E_{a}\right|\rho(t=0)\left|E_{b}\right\rangle \left\langle E_{b}\right.\left|n,l\right\rangle = \sum_{a,b}\exp\left(-i\omega_{ab}t\right)\phi_{mn}^{ab},\end{eqnarray*}

 where $\left|E_{a}\right\rangle $ and $\left|E_{b}\right\rangle $ are
energy eigenstates of the system under consideration, with energy values $E_{a}$ and $E_{b}$ respectively. In the last line, 
$\omega_{ab}$ is $(E_{a}-E_{b})/\hbar$ and \[\phi_{m,n}^{ab}=\sum_l\left\langle E_{a}\right|\rho(t=0)\left|E_{b}\right\rangle (\left\langle E_{b}\right|n,l\rangle \left\langle m,l\right|E_{a}\rangle), \] the last factor being nothing but the overlap between the 
$m$'th block of $\left|E_{a}\right\rangle $ and $n$'th block of
$\left|E_{b}\right\rangle $.

 Thus,
\begin{eqnarray}
Tr\left(\rho_{R}^{2}\right) & = & \sum_{m,n}\left(\left(\rho_{R}\right)_{m,n}\right)^{2}\nonumber \\
 & = & \sum_{a,b,a',b'}\left(\sum_{m,n}\phi_{m,n}^{ab}\phi_{m,n}^{a'b'}\right) \times\quad\exp\left(-i(\omega_{ab}+\omega_{a'b'})t\right).\label{eq:traceRho}\end{eqnarray}

 This shows that $Tr(\rho_{R}^{2})$ and hence the linear entropy
$S_{L}$must have Fourier components with frequencies depending upon
combinations of available energy intervals ($E_a-E_b$) (quasi-energies in
the case of time periodic systems such as the  kicked rotor). But the amplitudes 
of these Fourier components depend upon the overlap of energy eigenstates (
quasi-energy states for kicked systems ) over all available blocks
for the subsystems. Therefore, for Hamiltonians with a preponderance of large 
energy intervals and with eigenstates extended over all the basic
states, the power-spectra of $S_{L}$ will be broadband ones with considerably
large Fourier components for a wide range of frequencies. In the opposite case 
of Hamiltonians with a narrow range of available energy intervals, there will be only a relatively small number of components spread over a narrow frequency range. Evidently, this distinction will also show up in the temporal autocorrelations of the RDM elements, since the latter are simply the Fourier transforms of the corresponding power spectra. 

\subsubsection*{Available Energy Intervals}
\label{sec:AvailableEnergyIntervals}

It is known from the random matrix theory that the nearest neighbour level spacing
distribution (NNLSD) of energy spectra for a real symmetric random Hamiltonian
matrix is of the Wigner type, while it is of the Poisson type for a regular
Hamiltonian matrix.
\begin{figure}[h]
\centering
\includegraphics[height=6cm,width=9cm]{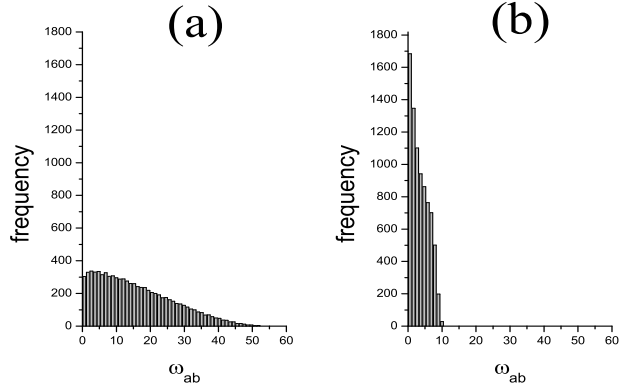}
\caption{\label{cap:EISpin}Distribution of all possible energy intervals
(divided by $\hbar$ ) for the Hamiltonian (a) $H_{c}$ and (b) $H_{r}$ (parameters as in Fig.~\ref{cap:AHspin}).}
\end{figure}
\begin{figure}[h]
\centering
\includegraphics[height=6cm,width=9cm]{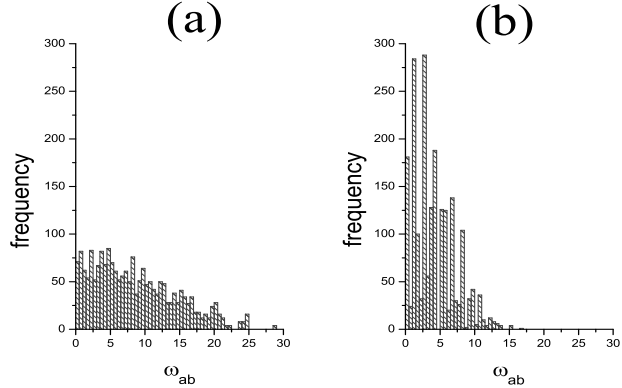}
\caption{\label{cap:EIHarper}Distribution of all possible energy intervals
(divided by $\hbar$ ) for the coupled Harper Hamiltonian for coupling
strengths as in Fig.\ref{cap:Poincare}.}
\end{figure}

However, for our  present purpose, the NNLSD is not of direct relevance, and of greater consequence is the distribution of  energy intervals between all pairs ($E_a, E_b$) determining the distribution of frequencies $\omega_{ab}=(E_{a}-E_{b})/\hbar.$ Since the available energy intervals are nothing but  additive combinations
of nearest neighbour level spacings, one expects that in accordance with the Wigner statistics, the distribution
of $\omega_{ab}$ will cover a wider range for a random Hamiltonian as compared to a  regular one. 

 This is corroborated in Fig.s \ref{cap:EISpin} and \ref{cap:EIHarper} for the spin systems (Sec.\ref{sec:TwoPrototypeSystems}) and the coupled Harper system (Sec.\ref{sec:CoupledHarperSystems}) where one finds that the energy intervals for a system with a greater degree of randomness in the Hamiltonian matrix elements are indeed spread over a larger interval as compared to the corresponding intervals for a regular Hamiltonian of similar type.

 In summary, the Wigner distribution of nearest level energy spacings, with its attendant distribution of available energy intervals, explains the dynamic features of quantum chaos as revealed through the RDM fluctuations.

\section{Hybrid Hamiltonians}
\label{sec:HybridHamiltonians}

 While the above can be looked at as a {\it prima facie} explanation of the characteristic features of RDM fluctuations, we present below further evidence of the role of energy eigenvalue and eigenvector distributions in generating these fluctuations, which may serve as a pointer towards a more detailed future explanation. For this, we consider a pair of contrived {\it hybrid} Hamiltonians with a view to explore separately the roles of eigenvalue and eigenvector distributions. We note that an arbitrary Hamiltonian $H$ may be written as, \begin{equation}
H=V^{-1}EV,\label{eq:eig-dist}\end{equation}
 where $V$ is the eigenvector matrix and E is a diagonal matrix with the 
eigenenergies as its diagonal elements.

 We first consider the spin systems discussed in Sec.\ref{sec:TwoPrototypeSystems}, as prototype instances of
regular and random systems. In accordance with Eq.\ref{eq:eig-dist}, $H_{r}$ and
$H_{c}$ may be written respectively as, 
\[
H_{r}=V_{r}^{-1}E_{r}V_{r}\]
and\[
H_{c}=V_{c}^{-1}E_{c}V_{c}.\]

 Using these, we construct the hybrid Hamiltonians
\begin{equation}
H_{rc}=V_{c}^{-1}E_{r}V_{c}\label{eq:Hrc}\end{equation}
and\begin{equation}
H_{cr}=V_{r}^{-1}E_{c}V_{r}.\label{eq:Hcr}\end{equation}
Evidently, $H_{rc}$ has the eigenenergies of the regular Hamiltonian $H_{r}$
and eigenvectors of the  chaotic Hamiltonian $H_{c}$. The case of
$H_{cr}$ is just the reverse, i.e., its energy eigenvalues are those of $H_c$ and eigenvectors are those of $H_r$ . We consider the time evolution for each of these two hybrid spin systems and compute the RDM for appropriately chosen subsystems, as in Sec.\ref{sec:TwoPrototypeSystems}.  
\begin{figure}[ht]
\centering
\includegraphics[height=6.5cm,width=10cm]{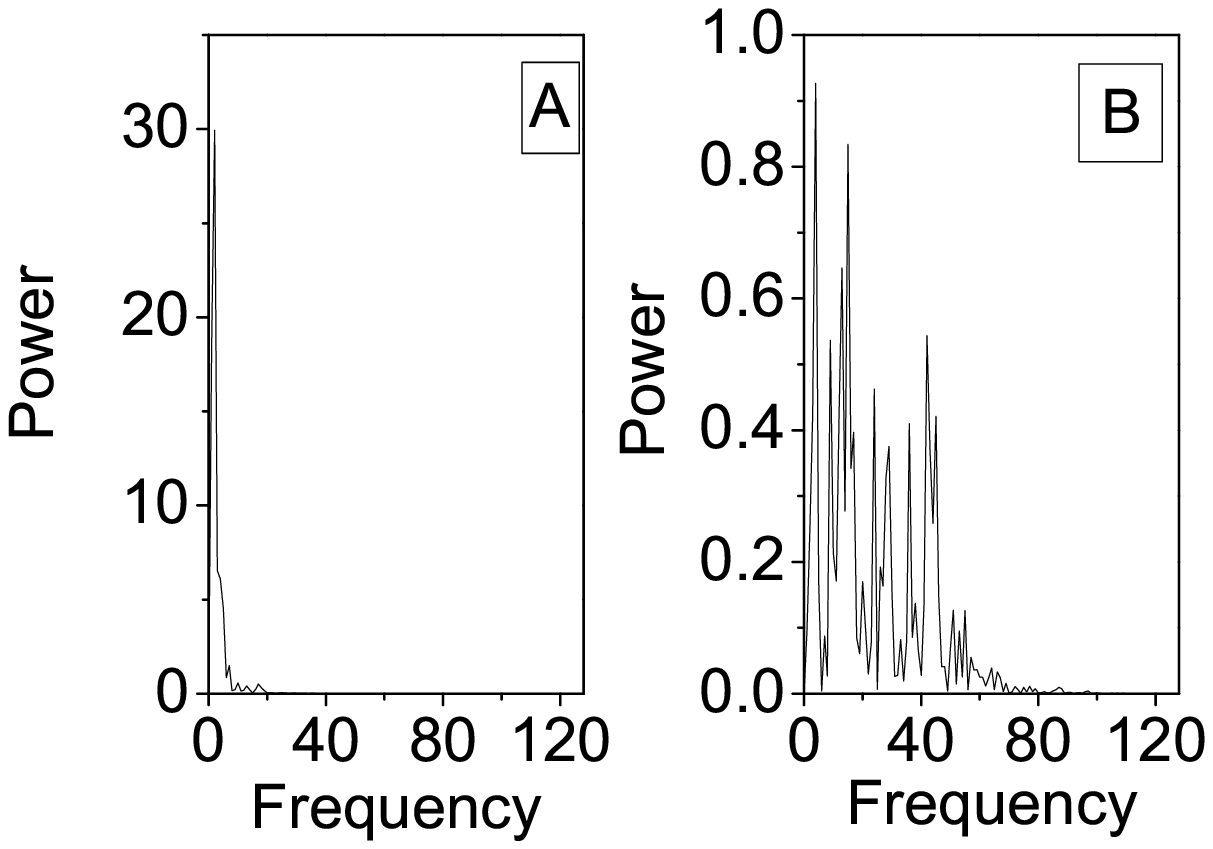}
\vskip 1cm
\centering
\includegraphics[height=6.5cm,width=10cm]{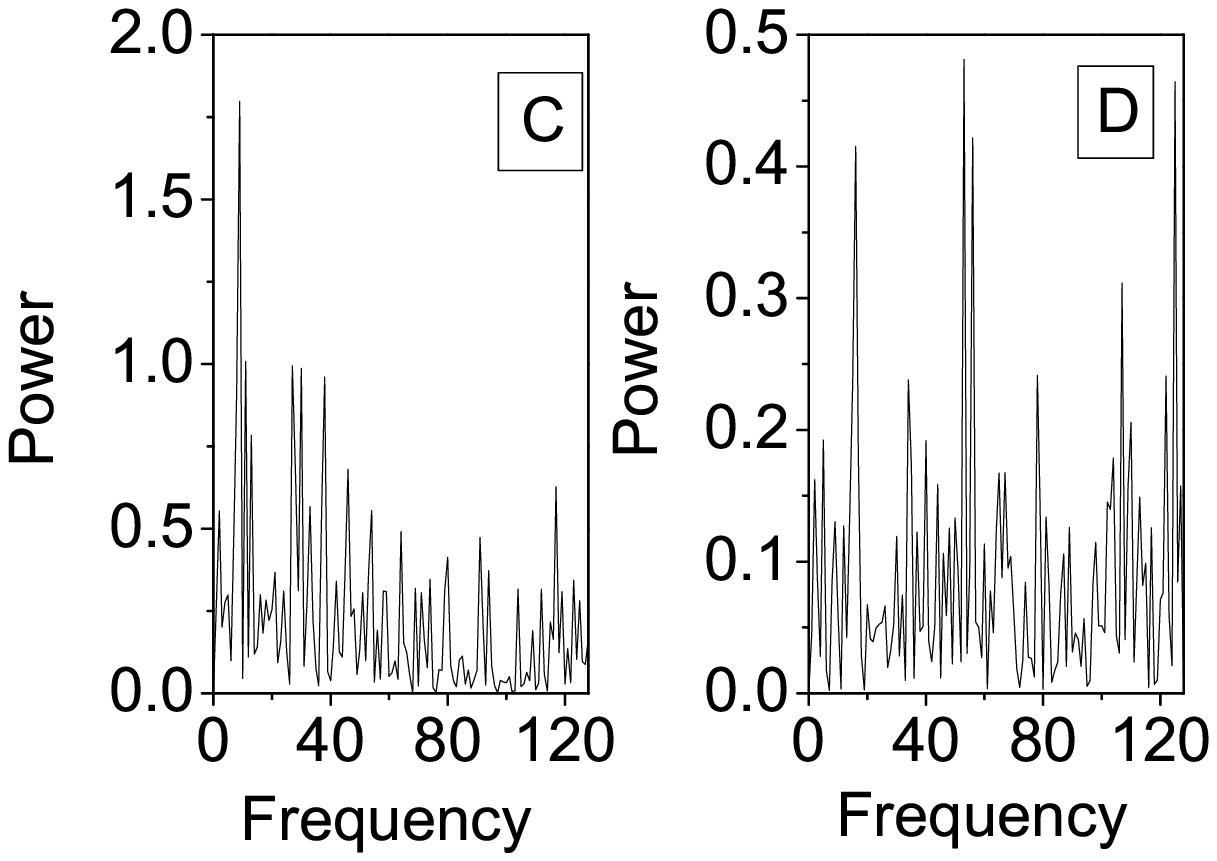}

\caption{\label{cap:mix}Power spetra of $S_L(=1-Tr(\rho_{R}^{2}))$ for spin systems when the Hamiltonian is (A) $H_{r}$, (B) $H_{rc}$, (C) $H_{cr}$ and (D) $H_{c}$ (see
text for explanation).}
\end{figure}

 Fig.\ref{cap:mix} shows the power spectra (reported earlier in ref.\cite{ncnsd} and reproduced here for the sake of completeness) of the time series for  $Tr\left(\rho_{R}^{2}\right)$ with Hamiltonians  $H_{rc}$ and $H_{cr}$, while those for $H_r$ and $H_c$ are also shown for comparison. 
 One observes that the broadening is slightly less
for $H_{cr}$ as compared to $H_{c}$. This is due to the fact that though
the available frequencies $\left(\omega_{ab}+\omega_{a'b'}\right)$
are the same for the two cases (since these depend on the available energy intervals 
alone), the amplitudes $\left(\phi_{mn}^{ab}\phi_{mn}^{a'b'}\right)$ (see Eq.\ref{eq:traceRho})
are small for most of the $m,n$'s due to the loacalised nature of the eigenvectors,
especially for higher frequencies. Again, for $H_{rc}$ and $H_{r}$ the available frequencies
cover a small range as compared to those for $H_{c}$ (see Sec.\ref{sec:AvailableEnergyIntervals}), resulting in 
narrower power spectra; still, the power spectrum is somewhat broader for $H_{rc}$ as compared to $H_{r}$ because of the extended nature of eigenvectors in $V_{c}$.

\section{Cocluding Remarks: the Classical Picture}
\label{sec:ComparisonWithClassicalEntropy}
While there exists a semiclassical analog of the von Neumann entropy (e.g., the Wherl entropy \cite{wehrl} defined in terms of the Husimi distribution) one cannot, strictly speaking, define a classical entropy in terms of the phase space distribution function $\rho(q,p).$ On the other hand, the concept of Kolmogoroff entropy applicable to a classically chaotic system does not have a direct analog for quantum systems. However, one can still seek an analogy between the time evolutions of the reduced density matrix for a quantum system and the reduced distribution function for a classical one (see below). 

 The main point we propose to make in the present paper, as in a couple of earlier papers, is that while dynamical features of classical chaos of a system are to be found in the time evolution of the system itself, analogous dynamical features for a quantum system are to be sought not in the system dynamics, but in the dynamics of {\it subsystems}. Thus, autocorrelations in the classical phase space of a chaotic system die down in the long run, while  corresponding autocorrelations in the Hilbert space vary quasiperiodically in the quantum case. On the other hand, autocorrelations for {\it subsystems} exhibit  similar dynamical features for the two situations, namely a decay in the long run. For a classically chaotic system, one infers this from the fact that that autocorrelations die down in the full phase space itself. For a quantum system, though the autocorrelations for the system itself do not die down in the long run, those for the subsystems do, as seen from the evidence presented above.  

 What happens in both the classical and quantum situations as one performs the reduction to a subsystem is, of course, a loss of reversibility. In the classical case, while the evolution of the distribution function in the full phase space is reversible and volume preserving, it is at the same time characterised by the feature of mixing. The individual points belonging to an ensemble are transported to far away regions of the phase space, developing complex patterns made up of whorls, tendrils and lacunae \cite{berry}, and this feature of mixing prevails on reduction to a subsystem, while at the same time there occurs a loss in reversibility.

 For a chaotic quantum system, on the other hand, it is the reduction itself that brings out the characteristic features of the system dynamics. The reduced density matrix is significantly different as compared to a regular system, and shows characteristic fluctuations of a random nature. While we have presented above a preliminary explanation of this feature, a more detailed analysis remains pending.

 We conclude with two remarks on possible future work. One relates to the formulation of criteria for quantum chaos {\it without reference to} the classical situation. Indeed, it seems desirable to have independent criteria for quantum chaos with a view to systems having no classical analogs (see \cite{AL} for an instance relating to the Baker's map), as also for characterising the dynamics of systems in the deep quantum regime where the semi-classical analysis bears no relevance~\cite{Allah}. The present approach, based on RDM fluctuations appears worthwhile from this point of view.

 The other observation one may be interested in relates to possible implications in statistical mechanics. Consider, for instance, the problem of approach to the canonical distribution, where the system $\bf{S}$ under consideration interacts weakly with a heat bath $\bf{H}$, making up a larger composite system $\bf{C}$. One looks at the evolution of $\bf{S}$, reducing from that of $\bf{C}$. The conventional approach is to go over to the limit of infinite number of degrees of freedom and a quasi-continuous spectrum for the heat bath $\bf{H}$ in order to arrive at the equilibrium distribution (see e.g. \cite{Allah}). An alternative approach would be to focus on the possible role of a chaotic interaction Hamiltonian between $\bf{S}$ and $\bf{H}$ in bringing about the equilibrium distribution for the reduced density matrix of $\bf{S}$. A preliminary presentation based on this point of view is in preparation \cite{AL-GG}.

\end{document}